\documentclass[11pt,a4paper]{article}
\usepackage{graphicx} 
\usepackage[utf8]{inputenc}
\usepackage{amsmath,amssymb,graphicx}
\usepackage{cite}
\usepackage{geometry}
\usepackage{tikz}
\usetikzlibrary{arrows.meta, positioning}
\usepackage{pgfplots}
\pgfplotsset{compat=1.18} 

\title{\textbf{Scattering of Squeezed Light by a Dielectric Slab}}

\author{\textbf{Ghafoor Pooseh}}
\date{December 18, 2025}

\makeatletter
\renewcommand{\maketitle}{
	\begin{center}
		{\LARGE \@title \par}
		\vspace{1.5em}
		{\large \@author \par}
		\vspace{0.5em}
		{\itshape Independent Researcher} \\
		\vspace{0.25em}
		{\ttfamily gpooseh@icloud.com} \par
		\vspace{1em}
		{\large \@date \par}
	\end{center}
	\vspace{1.5em}
}
\makeatother

\numberwithin{equation}{section}
\begin{document}
	
	\maketitle
	
	\begin{abstract}
		We develop a quantum theory for the scattering of squeezed coherent light by a dissipative dielectric slab. Using the Green-function quantization approach, we derive the transformation of the field quadratures and show how dispersion, absorption, and multiple reflections distort the incident squeezing. We find that the slab can selectively attenuate or amplify quadrature noise depending on the slab parameters and provide expressions for the output power spectra.
	\end{abstract}

	\section{Introduction}
	
	The transmission of classical light pulses through dispersive and absorbing media has long been studied in classical electrodynamics. In that picture, deformation of a pulse is explained by absorption, dispersion, and multiple reflections at interfaces. While this is accurate for macroscopic waves, it does not capture the behavior of light when it is treated as a quantum state. With the growth of quantum optics, and its applications in communication and precision measurement, it has become necessary to analyze how quantized states of light scatter and propagate in realistic media \cite{Loudon2000}.
	
	Earlier quantum optical studies investigated the transmission of coherent states through dissipative slabs \cite{Matloob1995,Gruner1996}. These works showed that only a full quantum treatment can describe how medium losses and fluctuations alter the noise properties of light. The Green--function approach to field quantization has been especially valuable here, since it includes both dispersion and absorption in a consistent way \cite{Matloob1996}.  
	
	A natural extension is to consider squeezed states of light. These states generalize coherent states by redistributing noise between quadratures \cite{Walls1983,Caves1981}. They are central resources for quantum communication, quantum information, and high--precision interferometry, since reduced fluctuations in one quadrature can push sensitivities below the shot--noise limit \cite{Slusher1985,Xiao1987,Vahlbruch2016}. This raises a fundamental and practical question: how do the dispersive and absorptive properties of a dielectric slab affect the squeezing engineered in the input light? Does the medium preserve, degrade, or reshape the squeezing?
	
	In an earlier work \cite{Pooseh2000}, we developed a quantum description of coherent light scattering by a dielectric slab using the Green--function quantization method. That analysis showed how the medium transforms coherent states and modifies their noise. The present study builds directly on that framework but extends it to squeezed coherent states, where quadrature fluctuations play a central role. In this sense, our work can be seen as a continuation and generalization of the coherent case.
	
	Other authors have studied related problems, such as propagation of squeezed light in linear dispersive systems and absorbing dielectrics \cite{Matloob1997,Khanbekyan2005,Knoll2001}, quantum noise in amplifying slabs \cite{Jeffers1993,Scheel1999}, and photon tunneling \cite{Tserkezis2011}. However, a full treatment of squeezed light scattering by a finite dielectric slab---in parallel with the coherent case---has not been presented.  
	
	In this paper, we develop a complete quantum theory of squeezed--light scattering by a dispersive and absorbing dielectric slab. We use the Green--function quantization framework together with input--output relations for boundary fields. After reviewing the operator formulation for slab geometry, we study single--mode squeezed coherent states and then extend the analysis to continuum squeezed states. In each case we examine quadrature variances, power spectra, and the average Poynting vector. The results identify conditions under which squeezing is preserved or altered, and clarify the roles of dispersion, absorption, and noise in realistic slab systems.
	Following a review of the field quantization formalism, we analyze single-mode squeezed state scattering, deriving expressions for the output quadrature variances. We then generalize this to continuum-mode squeezed pulses, examining their power spectrum and energy transport.

	\section{Field quantization}
	
\begin{figure}[h!]
		\centering
		\begin{tikzpicture}[scale=1.2, >=Stealth]
			
			\coordinate (O) at (0,0);
			\coordinate (Xmin) at (-5,0);
			\coordinate (Xmax) at (5,0);
			\coordinate (L) at (-2,0);
			\coordinate (R) at (2,0);

			\draw[->, thick] (Xmin) -- (Xmax) node[below] {$x$};

			\draw[thick] (-2,-1.5) node[below] {$x=-l$} -- ++(0,3) ;
			\draw[thick] (2,-1.5)  node[below] {$x=+l$}  -- ++(0,3) ;

			\fill[gray!20] (-2,1.5) rectangle (2,-1.5);

			\draw[->, thick, blue] (-3.5,0.8) -- (-2.5,0.8) node[midway,above] {$\hat{a}_{R1}$};
			\draw[<-, thick, red] (-3.5,-0.8) -- (-2.5,-0.8) node[midway,below] {$\hat{a}_{L1}$};
			
			\draw[->, thick, red] (2.5,0.8) -- (3.5,0.8) node[midway,above] {$\hat{a}_{R3}$};
			\draw[<-, thick, blue] (2.5,-0.8) -- (3.5,-0.8) node[midway,below] {$\hat{a}_{L3}$};
			
		\end{tikzpicture}
		\caption{Geometry of the dielectric slab and notation for the field operators. The slab has thickness $2l$ and is positioned between $x=-l$ and $x=+l$. Different domains are labeled 1-3. The dielectric function $\epsilon(x,\omega)$ changes value at the interfaces.}
		
		\label{fig:slab_geometry}
	\end{figure}
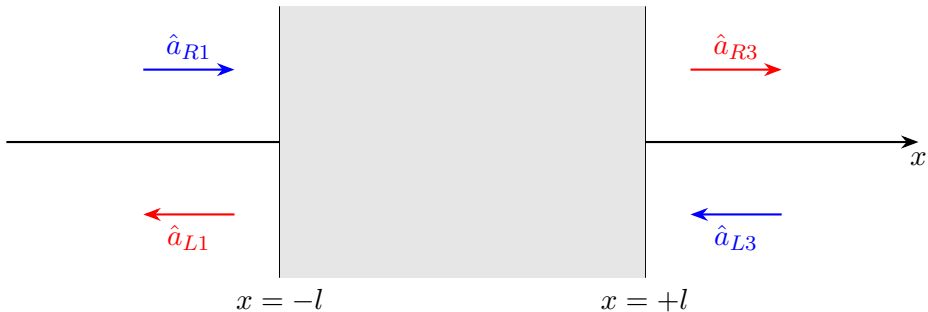

	To describe how light interacts with an absorbing and dispersive dielectric slab, we use the Green--function method of field quantization \cite{Matloob1995,Matloob1996,Gruner1996}. We consider a slab of thickness $2l$ with plane interfaces positioned at $x=-l$ and $x=+l$, as shown in Fig. ~\ref{fig:slab_geometry}. The dielectric function of the slab is assumed to be an arbitrary complex function of frequency, $\epsilon(\omega) = \epsilon_r(\omega) + i\epsilon_i(\omega)$, satisfying the Kramers--Kronig relations. The real and imaginary parts are related to the complex refractive index by
	\begin{equation}
	n^2(\omega) = \left[ \eta(\omega) + i\kappa(\omega) \right]^2 = \epsilon_r(\omega) + i\epsilon_i(\omega),
	\end{equation}
	where $\eta(\omega)$ and $\kappa(\omega)$ are the real refractive index and extinction coefficient, respectively. The complete spatial dependence is therefore
	\begin{equation}
	\epsilon(x,\omega) =
	\begin{cases}
		1, & |x| \geq l \\
		\epsilon(\omega), & |x| \leq l.
	\end{cases}
	\end{equation}
	It is useful to employ hereafter the label indices 1 through 3 for the domains $x<-l$, $-l<x<l$, and $x>l$, respectively.
	
	This approach is well suited for realistic materials, because it treats both dispersion and loss in a consistent way while keeping the basic commutation rules of quantum optics intact.
	
	\subsection{Field and noise operators}
	\textbf{Inside the slab}, working in frequency domain (Fourier convention \(f(t)=\int_0^\infty d\omega\, e^{-i\omega t} f(\omega)+\text{h.c.}\)), in macroscopic QED for absorbing media, positive frequency vector potential operator is written as:
	\begin{equation}
		\hat{\mathbf A}(\mathbf r,\omega)= \mu_0 \int d^3 r'\, \mathbf G(\mathbf r,\mathbf r',\omega)\, \hat{\mathbf j}_N(\mathbf r',\omega),
	\end{equation}
	where $\mathbf{G}(\mathbf{r},\mathbf{r}',\omega)$ is the classical Green tensor of the slab geometry. The slab contains microscopic charges (electrons, ions, or bound dipoles) that oscillate randomly due to thermal energy. These oscillations create local microscopic currents, represented by the noise current operator, $\hat{\mathbf{j}}_{\mathrm{N}}$, that represents absorption inside the medium. Within the framework of bosonic reservoir theory, the noise current operator is expressible in terms of the corresponding field operators~\cite{Matloob1995,Gruner1996,Matloob1997,Knoll2001,Scheel2008}.
	\begin{equation}
		\hat{\mathbf j}_N(\mathbf r,\omega)= \omega\sqrt{\tfrac{\hbar\,\varepsilon_0}{\pi}\,\varepsilon_I(\mathbf r,\omega)}\; \hat{\mathbf f}(\mathbf r,\omega).
	\end{equation}
	So, these noise sources are expressed through bosonic operators $\hat{f}(\mathbf{r},\omega)$, which obey
	\begin{equation}
		[\hat{f}(\mathbf{r},\omega),\hat{f}^{\dagger}(\mathbf{r}',\omega')] 
		= \delta(\mathbf{r}-\mathbf{r}')\,\delta(\omega-\omega'),
	\end{equation}
	ensuring that the electromagnetic field keeps the proper commutation relations even when losses are present. Projecting onto the left/right output channels gives effective noise operators

	\begin{subequations}
		\begin{equation}
		\hat F_A(\omega)= \int_{V_{\rm slab}} d^3 r\; \mathbf C_A(\mathbf r,\omega)\cdot \hat{\mathbf f}(\mathbf r,\omega), \qquad A\in\{L,R\},\label{eq:Xa}
		\end{equation}	
	with kernels $\mathbf C_A$ set by the Green tensor and boundary matching. Eq. (2.6) explicit the link between the microscopic bath operators $\hat{\mathbf f}(\mathbf r,\omega)$ 
	and the effective slab--port rightward an leftward noise operators $\hat F_{A}(\omega)$, and the noise state of the slab
	is represented by $|F\rangle$. The noise operators  have the following expectation values	
		\begin{equation}
			 \langle F |\hat F_A(\omega) |F\rangle = \langle F |\hat F_A^\dagger(\omega) |F\rangle=0 \qquad  A\in\{L,R\} \label{eq:Xb}
		\end{equation}
	\end{subequations}
	
	\textbf{Outside the slab}, the field is naturally described by right-- and left--moving plane--wave modes. On each side of the slab the positive frequency component of the vector potential operator can be written as
	\begin{equation}
		\hat{A}^+ (x,t) = \int_0^{+\infty} d\omega \left( \frac{\hbar}{4\pi \epsilon_0 c \omega \sigma} \right)^{1/2} \left[ \hat{a}_{R\Omega} (\omega) e^{i \omega x/c} + \hat{a}_{L\Omega} (\omega) e^{-i \omega x/c} \right] e^{-i \omega t},
	\end{equation}
	where $\Omega = 1,3$ refer to the different domain in question and $\sigma$ is the area of quantization in the $yz$ plane. The negative frequency part is obtained by taking the Hermitian conjugate of Eq. (2.7). We introduce operators $\hat{a}_{R}(\omega)$ and $\hat{a}_{L}(\omega)$ for these modes, with the commutator
	\begin{equation}
		[\hat{a}_{q}(\omega), \hat{a}_{q'}^{\dagger}(\omega')] = \delta_{qq'}\,\delta(\omega-\omega'),  \qquad q, q^\prime\in\{L,R\}
	\end{equation}
	These operators describe photons incident on and leaving the slab.
	The electric and magnetic fields are obtained from the quantized vector potential
	\begin{equation}
		\hat{E}(x,t) = -\partial \hat{A}(x,t)/\partial t, \quad \hat{B}(x,t) = \partial \hat{A}(x,t)/\partial x,
	\end{equation}
	where the particular choice of the gauge in which the scalar potential vanishes has been made.
	
	\subsection{Input-output relations and commutators}
	
	The slab couples the input and output fields in a linear way. In the presence of dissipation, it is convenient to use the concept of the modified scattering matrix to express the output modes in terms of the input modes
	\begin{equation}
		\begin{pmatrix}
			\hat{a}_{L1} (\omega) \\
			\hat{a}_{R3} (\omega)
		\end{pmatrix}
		=
		\begin{pmatrix}
			R_S (\omega) & T_S (\omega) \\
			T_S (\omega) & R_S (\omega)
		\end{pmatrix}
		\begin{pmatrix}
			\hat{a}_{R1} (\omega) \\
			\hat{a}_{L3} (\omega)
		\end{pmatrix}
		+ 
		\begin{pmatrix}
			\hat{F}_L (\omega) \\
			\hat{F}_R (\omega)
		\end{pmatrix},
	\end{equation}
	where the reflection and transmission coefficients of the slab are given by
	\begin{equation}
		R_S (\omega) = \frac{(n^2(\omega) - 1) \exp(-2i \omega l / c) [\exp(4i \omega n(\omega) l / c) - 1]}{(n(\omega) + 1)^2 - (n(\omega) - 1)^2 \exp(4i \omega n(\omega) l / c)},
	\end{equation}
	and
	\begin{equation}
		T_S(\omega) = \frac{4n(\omega)\exp[2i\omega(n(\omega)-1)l/c]}{(n(\omega)+1)^2-(n(\omega)-1)^2\exp(4i\omega n(\omega)l/c)}.
	\end{equation}
	
	The linear input--output relations (2.10) are in harmony with the quantum theory of lossy beam splitters \cite{Jeffers1993}. The operators $\hat{F}_L(\omega)$ and $\hat{F}_R(\omega)$ are associated with the noise contribution of the slab arising from its dissipative nature, These satisfy, by construction,
	\begin{equation}
	[\hat F_A(\omega),\hat F_B^\dagger(\omega')] = A(\omega)\,\delta_{AB}\,\delta(\omega-\omega'), \qquad A,B\in\{L,R\}
	\end{equation}
	where
	\begin{equation}
       A(\omega) \equiv 1 - |R_S(\omega)|^2 - |T_S(\omega)|^2
	\end{equation}
	is the absorptance (for a passive slab $A(\omega)\ge 0$). For a local thermal reservoir with occupation $n(\omega)$,
	\begin{equation}
       \langle \hat F_A^\dagger(\omega)\hat F_B(\omega')\rangle =  n(\omega)\,A(\omega)\,\delta_{AB}\,\delta(\omega-\omega'),
	\end{equation}
	The noise operators $\hat{F}_L(\omega)$ and $\hat{F}_R(\omega)$ are assumed to commute with the incoming mode operators of the slab.
	The latter operators themselves possess free field commutation relations:
	
	\begin{equation}
		\begin{aligned}
			\left[ \hat{a}_{R1}(\omega), \hat{a}^\dagger_{R1}(\omega') \right] &= \left[ \hat{a}_{L3}(\omega), \hat{a}^\dagger_{L3}(\omega') \right] = \delta(\omega - \omega'), \\
			\left[ \hat{a}_{R1}(\omega), \hat{a}^\dagger_{L3}(\omega') \right] &= \left[ \hat{a}_{L3}(\omega), \hat{a}^\dagger_{R1}(\omega') \right] = 0.
		\end{aligned}
	\end{equation}
	As the general property of an optical coupler with two input and two output ports, the output operators of the slab are independent whenever the input operators possess the free field commutation relation, therefore
	\begin{equation}
		\left[ \hat{a}_{L1}(\omega), \hat{a}^\dagger_{L1}(\omega') \right] = \left[ \hat{a}_{R3}(\omega), \hat{a}^\dagger_{R3}(\omega') \right] = \delta(\omega - \omega').
	\end{equation}
	This can be derived easily with the use of Eqs. (2.10) and (2.13).
	The present formalism is provided essentially for the continuum mode of the fields. However, as a general prescription \cite{Jeffers1993}, one may suppress the frequency dependence and replace the frequency delta function by unity to employ the latter formulation for discrete modes as well.
	
	This framework is the same as in our earlier study of coherent states \cite{Pooseh2000}. The new element here is the class of input states we consider. A squeezed coherent state is created by applying the squeezing operator
	\begin{equation}
		\hat{S}(\xi) = \exp\!\left[\tfrac{1}{2}(\xi^{*}\hat{a}^2 - \xi \hat{a}^{\dagger 2})\right],
	\end{equation}
	to the vacuum or to a coherent state. By combining this property with the input--output relations of the slab, we can track how quadrature noise and spectral properties of the squeezed field are changed by transmission and reflection.
	
	The operator framework developed in this section will be used in the following parts of the paper to analyze both single--mode and continuum squeezed states, and to identify how dispersion and absorption in the slab alter the observable squeezing.

\section{Single mode scattering}
A squeezed coherent state $|\alpha, \xi\rangle$ is defined by acting with the displacement operator $\hat{D}(\alpha)$ on the vacuum followed by the squeezing operator $\hat{S}(\xi)$
\begin{equation}
   |\alpha, \xi\rangle=\hat{S}(\xi) \hat{D}(\alpha)|0\rangle 
\end{equation}
where $\alpha$ and $\xi$ are two arbitrary complex parameters. The two operators $\hat{D}(\alpha)$ and $\hat{S}(\xi)$ are unitary transformations acting on the vacuum state of the electromagnetic field. The operator $\hat{D}(\alpha)$ displaces the vacuum in the complex $\alpha$-plane, while the operator $\hat{S}(\xi)$ squeezes the displaced vacuum. The squeezed coherent state can also be generated by first squeezing the vacuum, then displacing.

The squeezing operator possess the following properties:

\begin{equation}
    \hat{S}^{\dagger}(\xi) \hat{a} \hat{S}(\xi)  =\hat{a} \cosh \rho-\hat{a}^{\dagger} \exp (2 i \theta) \sinh \rho
\end{equation}

\begin{equation}
    \hat{S}^{\dagger}(\xi) \hat{a}^{\dagger} \hat{S}(\xi) =\hat{a}^{\dagger} \cosh \rho-\hat{a} \exp (-2 i \theta) \sinh \rho
\end{equation}
where $\xi=\rho \exp (2 i \theta)$. The action of the displacement operator on $\hat{a}$ and $\hat{a}^{\dagger}$ is of a form of a displacement, that is
\begin{equation}
    \hat{D}^{\dagger}(\alpha) \hat{a} \hat{D}(\alpha)  =\hat{a}+\alpha \\
\end{equation}

\begin{equation}
    \hat{D}^{\dagger}(\alpha) \hat{a}^{\dagger} \hat{D}(\alpha) =\hat{a}+\alpha^*
\end{equation}
The physical significance of the coherent state and squeezed coherent state are illuminated in the product of the uncertainties of the two field quadratures.

The linearly polarized monochromatic radiation field operator propagating along $x$-axis can be written in terms of the two Hermitian quadratures $\hat{X}$ and $\hat{Y}$ as
\begin{equation}
    \hat{E}(x, t)=E_0\{\hat{X} \cos (\omega t-k x)+\hat{Y} \sin (\omega t-k x)\},
\end{equation}
where $E_0$ is the electric field amplitude. The annihilation operator $\hat{a}$ is a linear combination of the two dimensionless operators $\hat{X}$ and $\hat{Y}$
\begin{equation}
    \hat{a}=\hat{X}+i \hat{Y} .
\end{equation}
The rotated complex amplitude is defined as
\begin{equation}
    \hat{X}^{\prime}+i \hat{Y}^{\prime}=(\hat{X}+i \hat{Y}) \exp (-i \theta)
\end{equation}
Using Eqs. (3.2) and (3.3), it is straightforward to show that
\begin{equation}
    \hat{S}^{\dagger}(\xi)\left(\hat{X}^{\prime}+i \hat{Y}^{\prime}\right) \hat{S}(\xi)=\hat{X}^{\prime} e^{-\rho}+i \hat{Y}^{\prime} e^\rho
\end{equation}
That is, the squeeze operator attenuates one of the rotated Hermitian quadratures amplitude and amplifies the other one. Employing the definition (3.1) along with Eqs. (3.2)-(3.5), one can easily show that the uncertainties of these two quadratures are
\begin{equation}
    \Delta X^{\prime}=\frac{1}{2} e^{-\rho}, \quad \Delta Y^{\prime}=\frac{1}{2} e^\rho
\end{equation}
and thus
\begin{equation}
    \Delta X^{\prime} \Delta Y^{\prime}=\frac{1}{4}
\end{equation}
There follows that the squeezed coherent states (3.1) display a class of minimum uncertainty states among them the coherent state with equal uncertainty in the both quadratures is a special case. In other words, the squeezed coherent states should necessarily possess reduced uncertainty in one quadrature, to be distinguished from the coherent state, at the expense of increased uncertainty in the other quadrature, to be satisfied in the minimum uncertainty relation (3.11).

Our considerations so far have been applied to the noise properties of the squeezed coherent light in general. However, in practical situations there will not be a case in which the transmission or reflection of light is not involved in one way or another. This will definitely affect on the latter noise properties. As was mentioned in the introduction, once scattering occurs, the interaction of the vacuum field with the incident light changes the all properties of the pulse in general and its noise properties in particular. Since it is the noise properties of the scattered light which is of special interest in the first place, we turn to this question here.

Consider that the incident rightward free field on the slab is a monochromatic squeezed coherent state and the incident leftward field is the vacuum field. The state of the system $|\psi\rangle$ is represented by
\begin{equation}
    |\psi\rangle=|\alpha, \xi\rangle_R|0\rangle_L|F\rangle
\end{equation}

The subscript indices " $R$ " and " $L$ " indicate to the direction of propagation. The rightward Hermitian quadratures in domain 3 are defined as
\begin{equation}
\hat{X}_{R 3}=\frac{1}{2}\left(\hat{a}_{R 3}+\hat{a}_{R 3}^{\dagger}\right)
\end{equation}
\begin{equation}
    \hat{Y}_{R 3}=\frac{1}{2 i}\left(\hat{a}_{R 3}-\hat{a}_{R 3}^{\dagger}\right)
\end{equation}
where $\hat{a}_{R 3}$ is given in terms of the incoming field modes of the slab by Eq. (2.10). The rotated complex amplitude can be written in the form of
\begin{equation}
    \hat{X}_{R 3}^{\prime}+i \hat{Y}_{R 3}^{\prime}=\left(\hat{X}_{R 3}+i \hat{Y}_{R 3}\right) \exp \left(-i \varphi_T\right)
\end{equation}
where $\varphi_T$ is the angle of rotation and will be given later. The subscript index " $T$ " denotes the fact that the rotation is associated with the transmitted light. Using Eqs. (2.10) and (2.16), one can easily show that
\begin{equation}
    \begin{array}{r}
\left\langle\left(\hat{X}_{R 3}^{\prime}\right)^2\right\rangle=\frac{1}{4}\left\{\left(T_S\right)^2 \exp \left(-2 i \varphi_T\right)\left\langle\left(\hat{a}_{R 1}\right)^2\right\rangle+\left(T_S^*\right)^2 \exp \left(2 i \varphi_T\right)\left\langle\left(\hat{a}_{R 3}^{\dagger}\right)^2\right\rangle\right. \\
\left.+1+2\left|T_S\right|^2\left\langle\hat{a}_{R 1}^{\dagger} \hat{a}_{R 1}\right\rangle+2\left\langle F_R^{\dagger} F_R\right\rangle\right\}
\end{array}
\end{equation}
and
\begin{equation}
    \left\langle\hat{X}_{R 3}^{\prime}\right\rangle^2=\frac{1}{4}\left\{T_S \exp \left(-i \varphi_T\right)\left\langle\hat{a}_{R 1}\right\rangle+T_S^* \exp \left(i \varphi_T\right)\left\langle\hat{a}_{R 1}^{\dagger}\right\rangle\right\}^2
\end{equation}
To simplify the later expressions it is advantageous to write the transmission coefficient of the slab in the form of

\begin{equation}
    T_S(\omega)=\left|T_S(\omega)\right| \exp \left(2 i \delta_T\right)
\end{equation}
The usual definition of the root-mean-square deviation along with the use of Eqs. (3.16) and (3.17) provide the uncertainty in this quadrature. Though rather lengthy, the calculations are straightforward and the result is
\begin{equation}
    \left(\Delta \hat{X}_{R 3}^{\prime}\right)^2=\frac{1}{4}\left\{1+2\left|T_S\right|^2\left[\sinh ^2 \rho-\sinh \rho \cosh \rho \cos 2\left[\varphi_T-\left(\theta+\delta_T\right)\right]\right]+2\left\langle F_R^{\dagger} F_R\right\rangle\right\}
\end{equation}
This gives the noise character of the rotated quadrature $\hat{X}_{R 3}^{\prime}$ as a function of $\varphi_T$, whose minimum is
 
\begin{equation}
    \left(\Delta \hat{X}_{R 3}^{\prime}\right)^2=\frac{1}{4}\left\{\left(1-\left|T_S\right|^2\right)+\left|T_S\right|^2 e^{-2 \rho}+2\left\langle F_R^{\dagger} F_R\right\rangle\right\}
\end{equation}
corresponding to the $\varphi_T=\theta+\delta_T$. We can easily show that the uncertainty in the other quadrature in this case is given by
\begin{equation}
    \left(\Delta \hat{Y}_{R 3}^{\prime}\right)^2=\frac{1}{4}\left\{\left(1-\left|T_S\right|^2\right)+\left|T_S\right|^2 e^{2 \rho}+2\left\langle F_R^{\dagger} F_R\right\rangle\right\}
\end{equation}
This can be derived by using Eqs. (2.10), (3.18) and the Hermitian conjugate of Eq. (3.15) for $\varphi_T=\theta+\delta_T$.

One may apply a similar treatment to the backward scattered light. However, the similarity between these two calculations allows us to write down the final result without giving the details. This is obtained by replacing the transmission coefficient with the reflection coefficient and changing the rightward noise term for the leftward counterpart in Eqs. (3.20) and (3.21). That is
\begin{equation}
  \left(\Delta \hat{X}_{L 1}^{\prime}\right)^2=\frac{1}{4}\left\{\left(1-\left|R_S\right|^2\right)+\left|R_S\right|^2 e^{-2 \rho}+2\left\langle F_L^{\dagger} F_L\right\rangle\right\}  
\end{equation}
and
\begin{equation}
    \left(\Delta \hat{Y}_{L 1}^{\prime}\right)^2=\frac{1}{4}\left\{\left(1-\left|R_S\right|^2\right)+\left|R_S\right|^2 e^{2 \rho}+2\left\langle F_L^{\dagger} F_L\right\rangle\right\}
\end{equation}
Note that in writing the latter expressions it is useful to consider the reflection coefficient of the slab of the form of
\begin{equation}
    R_S(\omega)=\left|R_S(\omega)\right| \exp \left(2 i \delta_R\right)
\end{equation}
The uncertainty in the rotated Hermitian amplitudes take their minimum values, given by Eqs. (3.22) and (3.23) for $\varphi_R=\theta+\delta_R$. The angle $\varphi_R$ represents the angle of rotation associated with the rotated backward complex amplitude.

As one expects, it is seen that the parameter $\alpha$ does not appear in Eqs. (3.20)-(3.23). In other words, it is only the squeezing factor of the incident light along with the characteristic parameters of the slab, including the noise property as well as the reflection and transmission coefficients, which establish the uncertainty of the different field quadratures of the scattered light. To illustrate the physical content of Eqs. (3.20)-(3.23) it is convenient to content ourselves in the first place with some special cases. Let us start with a non-dissipative or dissipative dispersive slab at zero temperature. Recalling Eqs. (2.10) and (2.15), this means that the noise contribution in expressions (3.20)-(3.23) must be set to zero. Therefore, for the forward scattered light we find that
\begin{equation}
    \left(\Delta \hat{X}_{R 3}^{\prime}\right)^2=\frac{1}{4}\left\{\left|R_S\right|^2+\left|T_S\right|^2 e^{-2 \rho}\right\}
\end{equation}
and
\begin{equation}
    \left(\Delta \hat{Y}_{R 3}^{\prime}\right)^2=\frac{1}{4}\left\{\left|R_S\right|^2+\left|T_S\right|^2 e^{2 \rho}\right\}
\end{equation}
The backward scattered light have the following uncertainties in this case
\begin{equation}
    \left(\Delta \hat{X}_{L 1}^{\prime}\right)^2=\frac{1}{4}\left\{\left|T_S\right|^2+\left|R_S\right|^2 e^{-2 \rho}\right\}
\end{equation}
and
\begin{equation}
    \left(\Delta \hat{Y}_{L 1}^{\prime}\right)^2=\frac{1}{4}\left\{\left|T_S\right|^2+\left|R_S\right|^2 e^{2 \rho}\right\}
\end{equation}
The product of the uncertainties of the two field quadratures for both the forward and backward scattered light is
\begin{equation}
   \left(\Delta \hat{X}_{R \Omega}^{\prime}\right)^2\left(\Delta \hat{Y}_{R \Omega}^{\prime}\right)^2=\frac{1}{16}\left\{1+\left|T_S R_S\right|^2(\cosh 2 \rho-1)\right\} 
\end{equation}
where $\Omega=1,3$. This expression can be derived by means of either Eqs. (3.25) and (3.26) or Eqs. (3.27) and (3.28).

If the transmitted light is of special concern, it then advantageous to keep the noise level of the forward scattered light as low as possible. This implies that
\begin{equation}
    \left|T_S(\omega)\right|=1, \quad \text { or } \quad\left|R_S(\omega)\right|=0
\end{equation}
The trivial solution of Eq. (3.30) is $T_S(\omega)=1$ and $R_S(\omega)=0$ for all frequencies. This evidently indicates to the propagation of the light in free space in the absence of the slab. In fact, due to the lack of any interaction between the leftward vacuum field and the incident light, the noise properties of the two field quadratures is not affected in this case. It is understood that the substitution of the latter solution into Eqs (3.27) and (3.28) provides the vacuum noise behavior for the both backward quadratures.

To look at a more practical situations, let us write Eq. (3.25) in the following form
\begin{equation}
    \left(\Delta \hat{X}_{R 3}^{\prime}\right)^2=\frac{1}{4}\left\{1-\left|T_S\right|^2\left(1-e^{-2 \rho}\right)\right\}
\end{equation}
This shows that the maxima of $\left|T_S\right|^2$ corresponds to the minima of $\Delta \hat{X}_{R 3}^{\prime}$ and vice versa. For a fixed value of the real refractive index of the slab we may choose either the frequency of the squeezed light for a particular $l$, or the width of the slab for a fixed $\omega$, so that to obtain the extrema of the uncertainty $\Delta \hat{X}_{R 3}^{\prime}$. This leads to the conditions
\begin{equation}
    \frac{4 \eta \omega l}{c}=2 m \pi
\end{equation}
for the minimum uncertainty and
\begin{equation}
    \frac{4 \eta \omega l}{c}=(2 m+1) \pi
\end{equation}
for the maximum of $\Delta \hat{X}_{R 3}^{\prime}$. Notice that the minimum noise behavior of the forward scattered light gives rise to the maximum uncertainty of the backward scattered light and vice versa. This is seen from Eqs. (3.25)-(3.28) easily. The condition (3.32) corresponds to the case where the forward scattered light is the same as the incident light, while the backward scattered light is the conventional vacuum. The imposition of Eq. (3.33) provides
\begin{equation}
    \left|T_S\right|=\frac{2 \eta}{\eta^2+1}
\end{equation}
and
\begin{equation}
    \left|R_S\right|=\frac{\eta^2-1}{\eta^2+1}
\end{equation}
Substitution of Eqs. (3.34) and (3.35) into Eqs. (3.25)-(3.28) shows that the transmitted and reflected lights are both squeezed, but none of them are coherent in this case. This demonstrates that for a lossless slab, perfect transmission (and thus perfect preservation of squeezing) is achievable at specific resonances, while away from these resonances, both transmitted and reflected light exhibit modified squeezing.

It is seen that the interaction of the vacuum field with the incident light is simply described by a unitary transformation for a non-dissipative slab. This brings a periodic behavior for the different quadratures of the scattered light as a function of $l$. We therefore have the occasion to keep the noise level of the transmitted light as low as the incident light by tuning the characteristic parameters of the slab with the frequency of the incident light. This corresponds to the destructive interferences that take place at $x=l$ for the reflected vacuum field. We may also improve the noise character of the reflected light if it is of special interest, but it may never retain the noise property of the incident light. In other words, despite the presence of the interaction of the incident light with the vacuum field, the wave nature of light allows us either to depress the effect for the transmitted light completely or to reduce the noise of the reflected light partially.

There is another special case for which Eqs. (3.20)-(3.23) are easily applicable with some care. Consider the propagation of a squeezed coherent light through a homogeneous absorbing medium. We first need some manipulation on the transmission and reflection coefficients of the slab to eliminate the presence of the different reflections and transmission which take place at the plane interfaces at $x= \pm l$. This is achieved by setting $t_1(\omega)=t_2(\omega)=1$ and $r(\omega)=0$ in Eqs. (2.11) and (2.12) for all frequencies. Therefore, apart from a phase factor which can be incorporated in the definition of the annihilation operator, we find that
\begin{equation}
    T_S(\omega) \rightarrow \exp [2 i \omega n(\omega) l / c], \quad R_S(\omega) \rightarrow 0
\end{equation}
Note that the substitution of the latter expressions into Eq.(2.10) provides the field modes before leaving the slab in terms of incoming modes inside the slab. In other words, this limit is only applicable when propagation over the length of $2 l$ in a homogeneous medium is concerned.

Substitution of Eq. (3.36) into Eqs. (3.20) and (3.21) gives the details of the noise properties of the transmitted light. It is seen that the propagation of the light in this case is accompanied with the increase of the uncertainty in the both quadratures. This increase depends on the extinction coefficient of the medium and the length $2 l$. The transmitted light is not, therefore, coherent but retains its squeezing character.

\begin{figure}[!h]
 	\includegraphics[width=\textwidth]{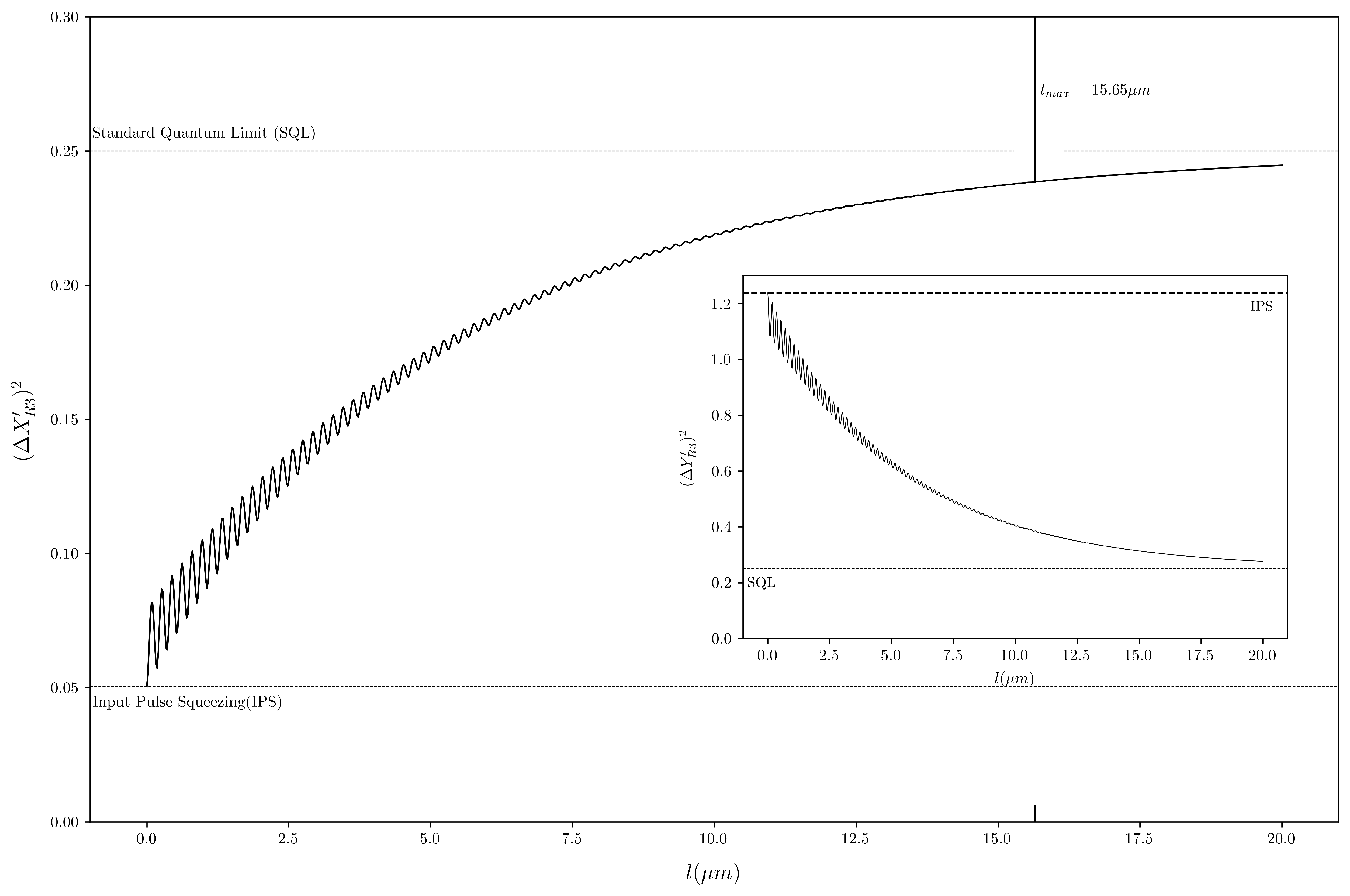}
 	\caption{
 		Variance of the transmitted squeezed quadrature ${\Delta \hat{X}_{R 3}^{\prime}}^2$ as a function of dielectric slab thickness for a squeezed coherent state with parameter$\rho = 0.8$ (initial variance = 0.045, equivalent to 6.9 dB of squeezing below the standard quantum limit). The calculation uses representative parameters for a dielectric material with refractive index $\eta = 1.5$ and extinction coefficient $\kappa=0.005$ at the squeezed light wavelength of $ \left (\lambda=1064 nm\right )$ nm. The oscillatory behavior with period $\lambda/(4\eta) = 0.177~\mu m$ results from quantum interference effects within the slab, with the variance approaching an asymptotic value as thickness increases. The inset shows the variance of the orthogonal quadrature  ${\Delta \hat{X}_{L 3}^{\prime}}^2$. The horizontal dashed gray line shows the original squeezed variance of the input pulse, and the dashed black line shows the standard quantum limit (SQL) or coherent state variance of $1/4$}.
 	\label{fig:DeltaXR3Variance_vs_Thickness}
 \end{figure}

Having these grounds, we now examine the noise properties of the scattered light due to presence of a dissipative dispersive slab. As in the case of non-absorbing slab we may wish to minimize the uncertainty $\Delta \hat{X}_{R 3}^{\prime}$ by tuning the characteristic parameters of the slab with the frequency of the incident light. One should note that the noise term of Eq. (3.20) is itself a function of the both $\omega$ and $l$. A numerical estimate of the different terms of this equation shows that for visible light at room temperature the order of the magnitude of this term is much smaller than the other terms. This term can, therefore, be eliminated in the evaluation of the extrema of $\Delta \hat{X}_{R 3}^{\prime}$ on assuming that the present formalism is restricted to either visible frequencies at room temperature or other ones at zero temperature. Expression (3.20), then, denote the fact that the maxima of $\left|T_S\right|^2$ practically corresponds to the minima of $\Delta \hat{X}_{R 3}^{\prime}$ and vice versa. The evaluation of these extrema, though straightforward, is rather lengthy. The frequency $\omega$ and the width of the slab should satisfy 
\begin{equation}
    \kappa B_\eta \sinh (4 \kappa \omega l / c)+\eta B_\kappa \sin (4 \eta \omega l / c)+2 \epsilon_i\left[A_{+} \cosh (4 \kappa \omega l / c)+A_{-} \cos (4 \eta \omega l / c)\right]=0
\end{equation}
where 
\begin{equation}
    A_{ \pm}=|\epsilon| \pm 1, \quad B_\eta=A_{+}^2+4 \eta^2, \quad B_\kappa=A_{-}^2-4 \kappa^2 .
\end{equation}
One can easily show that the solution of Eq. (3.37) in the limit of $\kappa \rightarrow 0$ is identical with Eqs. (3.32) and (3.33). Due to the presence of hyperbolic functions in Eq. (3.37), it is understood that the number of the different values of $l$ which satisfy this equation are finite. That is, for each values of $\eta$ and $\kappa$ there is a typical $l_{\max }$ beyond which the local extrema of $\Delta \hat{X}_{R 3}^{\prime}$ are evanescent due to the lossy behavior of the slab. We can show that $l_{\max}$ is given by
\begin{equation}
    l_{\max }=\frac{c}{4 \omega \kappa} \tanh ^{-1}\left\{\frac{-2 \epsilon_i \kappa A_{+} B_{\eta} \pm(O S C)_{\max }\left[D-4 \epsilon_i^2 A_{+}^2\right]^{1 / 2}}{D}\right\}
\end{equation}
where

\begin{equation}
    D=(O S C)_{\max }^2+\kappa^2 B_{\eta}^2
\end{equation}
in which $(O S C)_{\max }$ denotes the amplitude of the oscillatory part of Eq. (3.37). The solutions of Eq. (3.37) take a simple form if a slab of very poor absorption is concerned
\begin{equation}
    l=\frac{c}{4 \omega \eta}\left[\tan ^{-1}\left ( f\right) +m \pi\right], \quad m=1,2 \ldots
\end{equation}
with  
\begin{equation*}
f=\frac{-8\kappa \eta^2}{\left(\eta^2-1\right)^2}\quad \quad , \quad \quad 
f=\frac{+8\kappa}{\left(\eta^2-1\right)^2}
\end{equation*}
where only linear terms in $\kappa$ are kept. However, the evaluation of $l$ in terms of $\omega$ in the general case needs numerical calculations. The typical behavior of $\Delta \hat{X}_{R 3}^{\prime}$ is depicted in Fig.~\ref{fig:DeltaXR3Variance_vs_Thickness} as a function of $l$ for the representative values of the parameters. We see that the uncertainty of the squeezed quadrature of the forward scattered light has a very rapidly oscillatory behavior in terms of $l$, initiating from the value of the uncertainty of the squeezed quadrature of the incident light corresponding to $l=0$. The amplitude of these oscillations along with the squeezed behavior of this quadrature both reduces on increasing the width of the slab. As was mentioned, the oscillation is evanescent at $l_{\max }=15.65$   $\mu\mathrm{m}$ for the representative values of the parameters, but the output is always squeezed as far as the transmission coefficient differs from zero. The latter limit takes place for either $l \rightarrow \infty$ or $\kappa \rightarrow \infty$. The behavior of the other quadrature of the transmitted light is provided in the inset, which exhibits anti-squeezing behavior with variance values exceeding the standard quantum limit. These results demonstrate how the dielectric slab selectively modifies the quantum noise properties of squeezed light, preserving squeezing in one quadrature while enhancing noise in the orthogonal quadrature. 
 
The same considerations is applicable to the backward scattered light. As one sees from Eq. (3.23), it is the maxima of $\left|R_S\right|^2$ which corresponds to the minima of $\Delta \hat{X}_{L 1}^{\prime}$ in this case. For a fixed $\eta$ and $\kappa$, the extrema are obtained by the solutions of
\begin{equation}
    \begin{aligned}
& {\left[2 \kappa|\epsilon| \sinh (4 \omega \kappa l / c)+\epsilon_i \cosh (4 \omega \kappa l / c)\right] \cos (4 \omega \eta l / c)-\epsilon_i+} \\ \\
& \left[2 \eta|\epsilon| \cosh (4 \omega \kappa l / c)+\left(|\epsilon|^2+\epsilon_r\right) \sinh (4 \omega \kappa l / c)\right] \sin (4 \omega \eta l / c)=0
\end{aligned}
\end{equation}
It is seen that the solutions of this equation tend to Eq. (3.33) in the limit $\kappa \rightarrow 0$. One can easily show that for a very poor absorbing slab this equation can be rewritten of the form of
\begin{equation}
    \tan (2 \eta \omega l / c)=\left(\frac{\eta^2}{\kappa}\right)+\frac{2 \omega l}{c} \eta\left(\eta^2+1\right)
\end{equation}
\begin{figure}[tp!]
	\includegraphics[width=\textwidth]{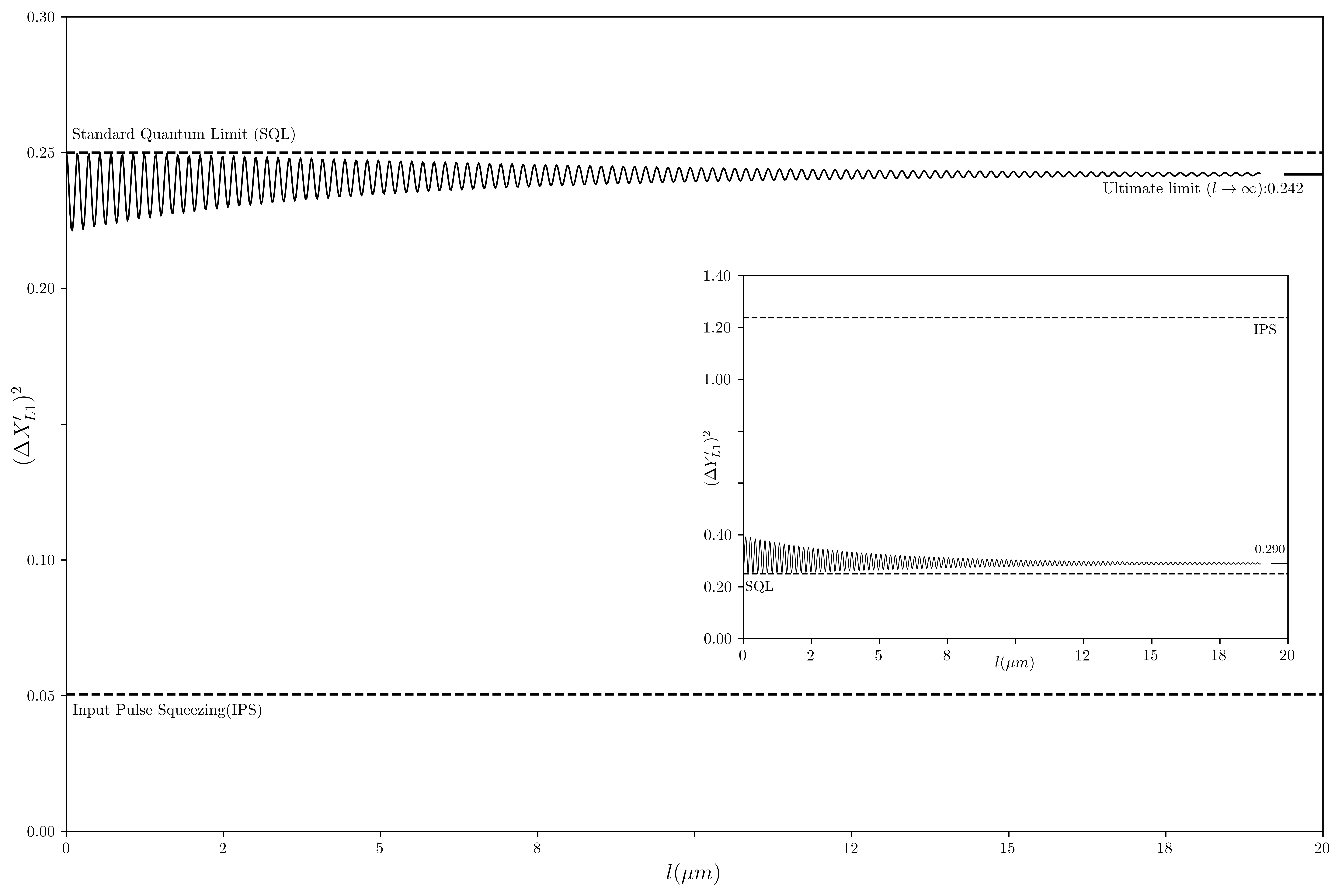}
	\caption{
		Variance of the reflected squeezed quadrature ${\Delta \hat{X}_{L1}^{\prime}}^2$ as a function of  slab thickness for a squeezed coherent state with parameter $\rho = 0.8$, $\eta  = 1.5$ and  $\kappa = 0.0075$ at the squeezed light wavelength of $ \lambda = 1064 nm$ .  The purple line shows the theoretical ultimate variance limit for an infinitely thick slab. The inset shows the variance of the orthogonal quadrature ${\Delta \hat{Y}_{L1}^{\prime}}^2$ for the reflected pulse, which exhibits anti-squeezing behavior with the variance approaching an ultimate limit of $0.290$  as thickness increases, above the standard quantum limit. }  
	\label{fig:DeltaXL1Variance_vs_Thickness}
\end{figure}
As for the forward scattering light, the general solutions of Eq. (3.42) are not obtained analytically. The variation of $\Delta \hat{X}_{L 1}^{\prime}$ in terms of $l$ is given in Fig.~\ref{fig:DeltaXL1Variance_vs_Thickness} for the same parameters as Fig.~\ref{fig:DeltaXR3Variance_vs_Thickness}. It is seen that the uncertainty of the squeezed quadrature possesses a very rapidly oscillating character in terms of $l$. This oscillatory behavior results from quantum interference effects within the slab, with the variance approaching an ultimate limit of $0.242$  as thickness increases. In contrary with the forward scattered light, the squeezing factor of the reflected light tends to zero when $l \rightarrow 0$ and eventually tends to a value less than that of the incident light on increasing the width of the slab. The imposition of the limit $l \rightarrow \infty$ on Eq. (3.22) yields
\begin{equation}
    \lim _{l \rightarrow \infty}\left(\Delta \hat{X}_{L 1}^{\prime}\right)^2=\frac{1}{4}\left\{1-\left(1-e^{-2 \rho}\right) \frac{\left|n^2-1\right|^2}{B_{\eta}+4 \eta A_{+}} \right\} 
\end{equation}
The inset of Fig.~\ref{fig:DeltaXL1Variance_vs_Thickness} shows the variation of $\left(\Delta \hat{Y}_{L 1}^{\prime}\right)^2$ in terms of $l$. As the plot shows this quadrature starts its oscillations from the vacuum value and tends to
\begin{equation}
    \lim _{l \rightarrow \infty}\left(\Delta \hat{Y}_{L 1}^{\prime}\right)^2=\frac{1}{4}\left\{1+\left(e^{2 \rho}-1\right) \frac{\left|n^2-1\right|^2}{B_{\eta}+4 \eta A_{+}}\right\}
\end{equation}
These results demonstrate how the dielectric slab modifies the quantum noise properties of reflected squeezed light, with different effects on the two orthogonal quadratures. As the final remark it should be noted that the present treatment can be generalized easily to the case in which the discrete multi-mode squeezed coherent light is concerned. However, the continuum mode of the squeezed coherent light needs a more careful examination.

\section{The continuum mode scattering}

The majority of optical devices generate the continuum mode squeezed coherent light. Such a pulse can be defined by acting the squeezing operator $\hat{S}[\xi(\omega)]$ on the continuum coherent state $|\{\alpha(\omega)\}\rangle$. For a continuum of modes, the squeezing operator generalizes from the single-mode case (Eq. 2.13) by integrating over all frequencies. The corresponding continuum-mode squeezing operator is:
\begin{equation}
    |\{\xi(\omega), \alpha(\omega)\}\rangle=\hat{S}[\xi(\omega)]|\{\alpha(\omega)\}\rangle .
\end{equation}
where the corresponding continuum-mode squeezing operator is:
\begin{equation}
	\hat{S}[\xi(\omega)] = \exp\left[\frac{1}{2} \int d\omega \left( \xi^*(\omega)\hat{a}^2(\omega) - \xi(\omega)\hat{a}^{\dagger 2}(\omega) \right) \right]
\end{equation}
This operator acts on the multi-mode vacuum state $|{0}\rangle$ to generate a continuum squeezed vacuum state.
Notice that the case in which the squeezing parameter $\xi$ is independent of $\omega$ is apparently the particular situation where the different frequency components of the continuum coherent state $\mid\{\alpha(\omega)\}\rangle $ are squeezed in the same way. As usual, we examine the case in which the incident rightwards and leftwards free field on the slab are the state given by Eq. (4.1) and the conventional vacuum respectively. The change in the frequency spread of the forward and backward scattered pulses are elucidated by considering the power spectra and average normal order Poynting vector.

\subsection{The power spectrum}

In analogy with the classical definition of the energy spectrum we adopt the following expression for the power spectrum $\mathcal{S}(\omega)$ in terms of the two-time correlation function of the field \cite{Cresser1983}
\begin{equation}
    \mathcal{S}(\omega)=\frac{1}{2 \pi} \int_{-\infty}^{+\infty} d t_1 \int_{-\infty}^{+\infty} d t_2 e^{-i \omega\left(t_1-t_2\right)}\left\langle\hat{E}^{-}\left(x, t_1\right) \hat{E}^{+}\left(x, t_2\right)\right\rangle
\end{equation}
This is apparently applicable to a non-stationary regime involving a pulse of light. The stationary counterpart of Eq. (4.3) is of the form of\cite{Cresser1983}
\begin{equation}
    \mathcal{S}(\omega)=\lim _{T \rightarrow \infty} \frac{1}{2 \pi T} \int_{-T / 2}^{T / 2} d t_1 \int_{-T / 2}^{T / 2} d t_2 e^{-i \omega\left(t_1-t_2\right)}\left\langle\hat{E}^{-}\left(x, t_1\right) \hat{E}^{+}\left(x, t_2\right)\right\rangle
\end{equation}
which will be used for the noise term. It turns out that these definition are directly related to the quantities observed in the light detection experiment.

Employing Eqs. (2.7) and (2.9), the positive frequency part of the electric field operator is
\begin{equation}
    \hat{E}^{+}(x, t)=i \int_0^{+\infty} d \omega\left(\frac{\hbar \omega}{4 \pi \epsilon_0 c \sigma}\right)^{1 / 2}\left[\hat{a}_{R \Omega}(\omega) e^{i \omega x / c}+\hat{a}_{L \Omega}(\omega) e^{-i \omega x / c}\right] e^{-i \omega t}
\end{equation}
The negative frequency part is determined by taking the Hermitian conjugate of Eq. (4.5). The general state of the system is
\begin{equation}
    |\{\psi(\omega)\}\rangle=|\{\xi(\omega), \alpha(\omega)\}\rangle_R|0\rangle_L|F\rangle
\end{equation}
To clarify the calculation it is convenient in the first place to restrict our discussion to a slab at zero temperature. Substitution of Eq. (4.5) with $\Omega=3$ and its negative frequency counterpart into Eq. (4.3) yields
\begin{equation}
    \mathcal{S}_T(\omega)=\left(\frac{\hbar \omega}{2 \epsilon_0 c \sigma}\right)\langle\{\psi(\omega)\}| \hat{a}_{R 3}^{\dagger}(\omega) \hat{a}_{R 3}(\omega)|\{\psi(\omega)\}\rangle,
\end{equation}
where Eq. \eqref{eq:Xb} and the vanishing of terms in which the annihilation operator acts directly on the vacuum are applied. We use henceforth the subscript indices " $I$ ", " $T$ ", " $R$ " and " $N$ " to designate the power spectrum and Poynting vector for incident, transmitted, reflected and noise label respectively. The rightwards annihilation and creation operators in domain 3 can be expressed in terms of the input mode operators and $\hat{F}_R(\omega)$ by means of Eq. (2.10). Applying these expressions, after some algebra, Eq. (4.7) takes the simple form
\begin{equation}
    \mathcal{S}_T(\omega)=\left|T_S(\omega)\right|^2 \mathcal{S}_I(\omega)
\end{equation}
where the incident spectrum is
\begin{equation}
    \mathcal{S}_I(\omega)=\left(\frac{\hbar \omega}{2 \epsilon_0 c \sigma}\right)|\alpha(\omega)|^2\{\cosh [2 \rho(\omega)]-\sinh [2 \rho(\omega)] \cos (2 \theta(\omega)-2 \varphi(\omega))\}
\end{equation}
in which $\alpha(\omega)=|\alpha(\omega)| \exp [i \varphi(\omega)]$ and $\xi(\omega)=\rho(\omega) \exp [2 i \theta(\omega)]$. A similar treatment for the backward scattering show that
\begin{equation}
    \mathcal{S}_R(\omega)=\left|R_S(\omega)\right|^2 \mathcal{S}_I(\omega)
\end{equation}
Note that for a dissipative dispersive slab in thermal equilibrium at finite temperature $T$, an additional term appears in Eq. (4.7):
\begin{equation}
	\mathcal{S}_T(\omega)=\left(\frac{\hbar \omega}{2 \epsilon_0 c \sigma}\right)\left[\langle\{\psi(\omega)\}| \hat{a}_{R 3}^{\dagger}(\omega) \hat{a}_{R 3}(\omega)|\{\psi(\omega)\}\rangle + \langle F| \hat F_R^\dagger(\omega)\hat F_R(\omega)|F\rangle\right],
\end{equation}
A term independent of pulse characteristic of the form
\begin{equation}
    \mathcal{S}_N(\omega)=\left(\frac{\hbar \omega}{2 \epsilon_0 c \sigma}\right) \bar{n}(\omega, T) A\left(\omega\right), \qquad  A(\omega) \equiv 1 - |R_S(\omega)|^2 - |T_S(\omega)|^2
\end{equation}
must be added to Eqs. (4.8) and (4.10). This term has been achieved by using Eq. (4.4) for the stationary radiation of the slab and Eq. (2.15). $A\left(\omega\right)$ is the \textit{absorptance} of slab and,

\begin{equation}
    \bar{n}(\omega) = \frac{1}{\exp\left(\frac{\hbar \omega}{ k_B T}\right) - 1}
\end{equation}
is the Bose--Einstein occupation number of the thermal photons in which $\hbar=h/2\pi$, h is Planck’s constant and $k_B$ is the Boltzmann factor.

The existence of the slab transmission and reflection coefficients in Eqs. (4.8) and (4.10) means that the power spectra of the scattered pulses, compared to the incident pulse, are changed. To examine these two expressions. it is advantageous to illuminate the physical content of the incident spectrum first. This is achieved by assuming either $(\theta-\varphi)=0$ or $(\theta-\varphi)=\pi / 2$ in Eq. (4.9). One can easily show that the latter leads to
\begin{equation}
    \mathcal{S}_I(\omega)=\left(\frac{\hbar \omega}{2 \epsilon_0 c \sigma}\right)|\alpha(\omega)|^2 \exp [2 \rho(\omega)]
\end{equation}
while the former is associated with
\begin{equation}
    \mathcal{S}_I(\omega)=\left(\frac{\hbar \omega}{2 \epsilon_0 c \sigma}\right)|\alpha(\omega)|^2 \exp [-2 \rho(\omega)]
\end{equation}


It is seen that for $\rho \rightarrow \infty$ the spectrum given by Eq. (4.12) tends to zero, and thus corresponds to the spectrum of the squeezed quadrature of the incident field. The expression (4.14) coincide with the spectrum of the other quadrature, since it tends to zero when $\rho \rightarrow-\infty$. It is clear that in the theory of scattering we are mainly interested to study the behavior of the spectrum of the squeezed quadrature, given by Eq. (4.15), rather than Eq. (4.14).

It is helpful to focus our considerations on the case in which $a(\omega) \rightarrow 1$. Note that it is not only a matter of simplification which motivates us to examine this limit first. Under certain conditions the output of a degenerate parametric amplifier is approximated by a single continuous mode squeezed vacuum \cite{Blow1990}. The spectrum of the incident squeezed quadrature in this regime is
\begin{equation}
    \mathcal{S}_I(\omega)=\left(\frac{\hbar \omega}{2 \epsilon_0 c \sigma}\right) \exp [-2 \rho(\omega)]
\end{equation}
and the spectra of the scattered pulses in this case are
\begin{equation}
\mathcal{S}_{\Gamma}(\omega)=\left(\frac{\hbar \omega}{2 \epsilon_0 c \sigma}\right)\left|C_{\Gamma}(\omega)\right|^2 \exp [-2 \rho(\omega)], \qquad \Gamma\in \{T,R\}
\end{equation}
where  
\begin{equation}
    C_{\Gamma}(\omega)= \begin{cases}T_S(\omega), & \Gamma=T \\ R_S(\omega), & \Gamma=R\end{cases}
\end{equation}
Comparing Eqs. (4.15) and Eq. (4.17), it is seen that the spectrum of scattered squeezed vacuum possesses the same form as the spectrum of the incident squeezed coherent light provided that $C_{\Gamma}(\omega) \rightarrow \alpha(\omega)$. Similarly, the spectrum of the scattered squeezed coherent light has the same form as the incident pulse if $C_{\Gamma}(\omega) \alpha(\omega) \rightarrow \alpha(\omega)$. Therefore, the spectra of the scattered light in both cases are easily obtained if we examine the spectrum of the form Eq. (4.17) and comparing the result with the spectrum given by Eq. (4.16).

A squeezed pulse is generally described by a frequency distribution function of the form of a wave packet. Consider a Gaussian distribution centered on $\omega_c$. that is
\begin{equation}
    \rho_I(\omega)=\rho_{I} \exp \left[-\mathcal{L}_I^2\left(\omega-\omega_c\right)^2 / 4 c^2\right]
\end{equation}
where $\mathcal{L}_I^2$ is the mean-square spatial length of the pulse and $\rho_I$ determines the maximum amount of squeezing at the central frequency. 
It is convenient to add subscript $I$ to  $\rho(\omega)$ to explicitly denote that it corresponds to the incident pulse. 
It is customary to choose the frequency spread of the packet to much smaller than the central frequency
\begin{equation}
    c / \mathcal{L}_I \ll \omega_c
\end{equation}
It is also convenient to use the narrow-band-width approximation for the incident pulse which is usually applicable in practical situations. In this limit the variation of the slab dielectric function as a function of frequency is taken to be smooth enough and furthermore, the band-width of the incident pulse is assumed to be so narrow that the variation of the real refractive index $\eta(\omega)$ and extinction coefficient $\kappa(\omega)$ as a function of frequency are such that the complex optical wave vector can be expanded around $\omega=\omega_c$, keeping only the first two terms of the expansion \cite{Born1999,Artoni1997}. Furthermore, to retain the form of a single pulse for the output let us assume that the pulse length is much greater than the optical thickness of the slab. that is
\begin{equation}
 \quad \mathcal{L}_I \gg 2 l \eta_c
\end{equation}
in the case that pulse length is much smaller than optical thickness of the slab, there will be trains of pulses for both transmitted and reflected light whose peak values decrease exponentially due to the presence of dissipation inside the slab. In the intermediate domain, we will have  forward and backward outputs as trains of pulses whose separations are smaller than the pulse lengths and thus the interference terms have appreciable effect[\cite{Pooseh2000}]. Expanding Eq. (4.19) around $\omega=\omega_c$ and keeping only the first two non vanishing terms of the expansion, we find that
\begin{equation}
    \rho_{I}(\omega)=\rho_I\left[1-\mathcal{L}_I^2\left(\omega-\omega_c\right)^2 / 4 c^2\right]
\end{equation}
The transmission and reflection coefficients of the slab, subjected to the narrow-band-width approximation, can be written as \cite{Pooseh2000}
\begin{equation}
    C_{\Gamma}(\omega)=C_{\Gamma}\left(\omega_c\right) \exp \left[2 i \gamma_{\Gamma} l\left(\omega-\omega_c\right)-8 \beta_{\Gamma}^2 l^2\left(\omega-\omega_c\right)^2\right]
\end{equation}
The different parameters appearing in Eq. (4.23) are defined as
\begin{equation}
\gamma_T=k_c^{\prime}-\frac{1}{c}+\frac{2 k_c^{\prime}\left(n_c-1\right)^2 \exp \left(4 i \omega_c n_c l / c\right)}{\left(n_c+1\right)^2-\left(n_c-1\right)^2 \exp \left(4 i \omega_c n_c l / c\right)} 
\end{equation}
\begin{equation}
    \beta_T=\frac{k_c^{\prime}\left(n_c^2-1\right) \exp \left(2 i \omega_c n_c l / c\right)}{\left(n_c+1\right)^2-\left(n_c-1\right)^2 \exp \left(4 i \omega_c n_c l / c\right)}
\end{equation}
and
\begin{equation}
\gamma_R=\gamma_T+k_c^{\prime}\left(\frac{\exp \left(4 i \omega_c n_c l / c\right)+1}{\exp \left(4 i \omega_c n_c l / c\right)-1}\right)
 \end{equation}
\begin{equation}
    \beta_R^2=\beta_T^2-k_c^{\prime 2} \frac{\exp \left(4 i \omega_c n_c l / c\right)}{\left[\exp \left(4 i \omega_c n_c l / c\right)-1\right]^2}
\end{equation}
The prime denotes the frequency derivative and the subscript index " $c$ " of each function shows that it is evaluated at the central frequency. As in Ref. \cite{Pooseh2000}, it is convenient to use
\begin{equation}
k_{c r}^{\prime} \equiv \frac{1}{v_g} \simeq \frac{\eta_c}{c}
\end{equation}
\begin{equation}
    k_{c i}^{\prime} \simeq \frac{\kappa_c}{c}
\end{equation}
for the numerical calculations.

Using Eq. (4.23), one can easily show that
\begin{equation}
    \left|C_{\Gamma}(\omega)\right|=\exp \left[\ln \left|C_{\Gamma}\left(\omega_c\right)\right|-2 \operatorname{Im}\left(\gamma_{\Gamma}\right) l \left(\omega-\omega_c\right)-8 \operatorname{Re}\left(\beta_{\Gamma}^2\right) l^2 \left(\omega-\omega_c\right)^2\right]
\end{equation}
This is obtained by expanding $\left|C_{\Gamma}(\omega)\right|$ around $\omega=\omega_c$ and keeping only the first three terms. Combining Eqs. (4.17), (4.22) and (4.30), the spectrum of the scattered squeezed vacuum can be rewritten in the form of
\begin{equation}
    \mathcal{S}_{\Gamma}(\omega)=\left(\frac{\hbar \omega}{2 \epsilon_0 c \sigma}\right) \exp \left\{-2 \rho_{\Gamma}\left[1-\frac{\mathcal{L}_{\Gamma}^2\left(\omega-\omega_c+\Delta \omega_{\Gamma}\right)^2}{4 c^2}\right]\right\}
\end{equation}
where the frequency shift $\Delta \omega_{\Gamma}$, the mean square length of the pulse $\mathcal{L}_{\Gamma}^2$ and the squeezing parameter $\rho_{\Gamma}$ are
\begin{equation}
\Delta \omega_{\Gamma}=-\operatorname{Im}\left(\gamma_{\Gamma}\right)l \left[\frac{\rho_I \mathcal{L}_{I}^2}{4 c^2}-8 \operatorname{Re}\left(\beta_{\Gamma}^2\right) l^2 \right]^{-1}
\end{equation}
\begin{equation}
    \rho_{\Gamma}=\left[\rho_I-\ln \left|C_{\Gamma}\left(\omega_c\right)\right|\right]-\operatorname{Im}\left(\gamma_{\Gamma}\right) l \Delta \omega_{\Gamma}
\end{equation}
and
\begin{equation}
    \frac{\mathcal{L}_{\Gamma}^2}{4 c^2}=-\frac{\operatorname{Im}\left(\gamma_{\Gamma}\right) l }{\rho_{\Gamma} \Delta \omega_{\Gamma}}
\end{equation}
In order to compare this spectrum with the spectrum given by Eq. (4.16), it is convenient to rephrase Eq. (4.31) in the following form
\begin{equation}
    \mathcal{S}_{\Gamma}(\omega)=\left(\frac{\hbar \omega}{2 \epsilon_0 c \sigma}\right) \exp \left[-2 \rho_{\Gamma}(\omega)\right]
\end{equation}
where
\begin{equation}
    \rho_{\Gamma}(\omega)= \rho_{\Gamma}\left[1-\frac{\mathcal{L}_{\Gamma}^2\left(\omega-\omega_c+\Delta \omega_{\Gamma}\right)^2}{4 c^2}\right]
\end{equation}
By comparing Eqs. (4.36) and (4.22), one can easily see the effect of scattering  by a dielectric slab on squeezing spectrum of incident pulse. It is seen that the forward and backward pulses are squeezed having Gaussian distributions, centered on $\omega=\omega_c-\Delta \omega_{\Gamma}$. So, we have a shift in frequency of squeezing  distribution peak point. Also, the mean square length of both transmitted and reflected pulses has  exhibited a variation relative to that of the input pulse. It is good to remember, all these equations are valid only under approximations that took place to extracting them. Also, interpreting some results parameters as a real physical quantity could be misleading sometimes. As an example, seemingly, $\rho_{\Gamma}$ in Eq. (4.36) is the real squeezing factor of scattered pulses in their peak frequencies; If one evaluate or plot $\rho_{\Gamma}/\rho_{I}$, one will see that the values all are greater than 1, that means squeezing in both transmitted and reflected pulses are better than incident one; an obvious violation of fundamental rules. In fact, because of the dependence of second term of Eq. (4.36), the term in bracket, for any values of frequencies or real refractive indices, decrease $\rho_{\Gamma}(\omega)$ such that, its value is always smaller than $\rho_{I}(\omega)$. One can defines real physical effective squeezing factor as:
\begin{equation}
	\rho_{\Gamma}^{\prime}=\rho_{\Gamma}\left[1-\frac{\mathcal{L}_{\Gamma}^2\left(\Delta \omega_{\Gamma}\right)^2}{4 c^2}\right] \frac{\mathcal{L}_{\Gamma}^2}{\mathcal{L}_I^2}
\end{equation}
that is always smaller than $\rho_{I}$. 
Another similar source of misleading is the asymptotic behavior of quantities such as, $\Delta\omega_{\Gamma}$, $\Delta\mathcal{L}_{\Gamma}$ and other similar quantities. In almost all equations, we have an oscillatory term that its oscillation frequency is depends on all $l$, $\omega$ and $\eta$, when the denominator gets close to zero (narrow bandwidth approximation breaking down), the ratio can blow up, giving unreal values much larger than 1. To validate the behavior of the reflection and transmission contributions, we evaluated the exact scattering amplitudes at the central frequency using high-precision numerics. For $\eta\rightarrow 1$, the transmission coefficient approaches unity and the reflection coefficient vanishes as $\kappa \rightarrow 0$. In fact, the energy-weighted reflected quantity $Q_R$ scales as $\kappa^2$, while the transmitted fraction $Q_T$ approaches one. This confirms that the apparent anomaly observed in the asymptotic expansion disappears when the exact formulas are used, ensuring consistency with energy conservation and the physical expectation that a lossless matched slab must transmit fully. Regions where values of these quantities are large mark a breakdown of the narrow-band assumption and should be avoided for faithful pulse scattering analysis; conversely, they may be exploited for sensitive refractive-index sensing provided loss and stability are controlled.

\begin{figure}[!ht]
  	\includegraphics[width=\textwidth]{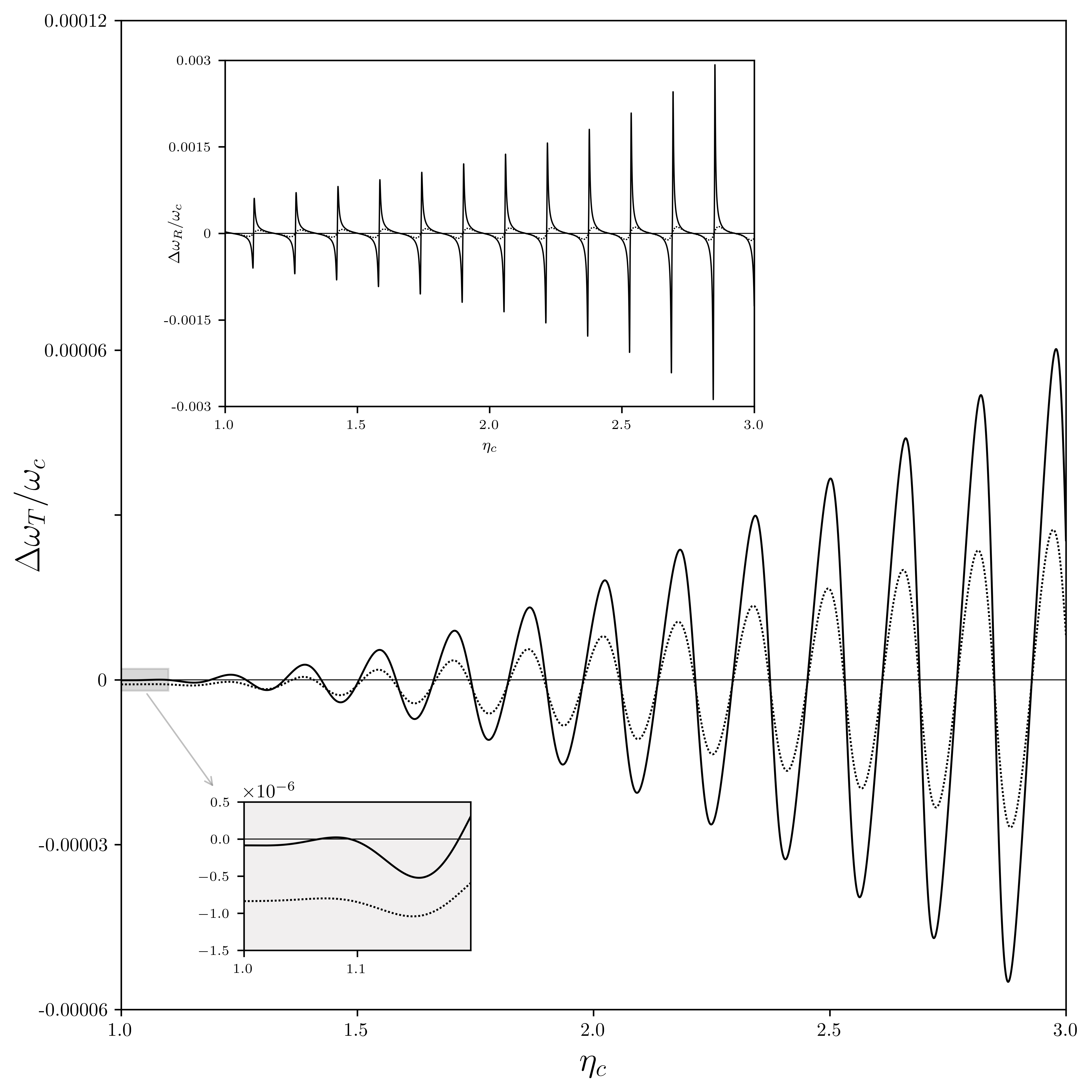}
  	\caption{Relative frequency shift, 
  		     $\Delta\omega_{\Gamma}/ \omega_c$, 
  			 for transmitted (main) and reflected (inset) squeezed pulses as a function of slab refractive index,$\eta_c$, using representative parameters($\lambda_c=633$ $nm$, $\mathcal{L}_I / l =80$, and $\rho_{I}=1.5$) for two values of extinction coefficient, $\kappa_c$(solid for $\kappa_c$=0.002, and dotted for $\kappa_c$=0.02).  
  		     The oscillatory structure arises from Fabry-Perot resonances within the slab. Increased loss (`$\kappa_c$=0.02`, dotted) dampens these resonances compared to the lower-loss case (`$\kappa_c$=0.002`, solid). The shift vanishes for a matched, lossless slab (`$\eta_c$→1`, `$\kappa_c$→0`), All oscillating pattern shifts downward as $\kappa$ increases, as shown in the detailed inset(lower left).}
  	\label{fig:Relative_Delta_omega_vs_eta}
\end{figure}

\begin{figure}[!ht]
  	\includegraphics[width=\textwidth]{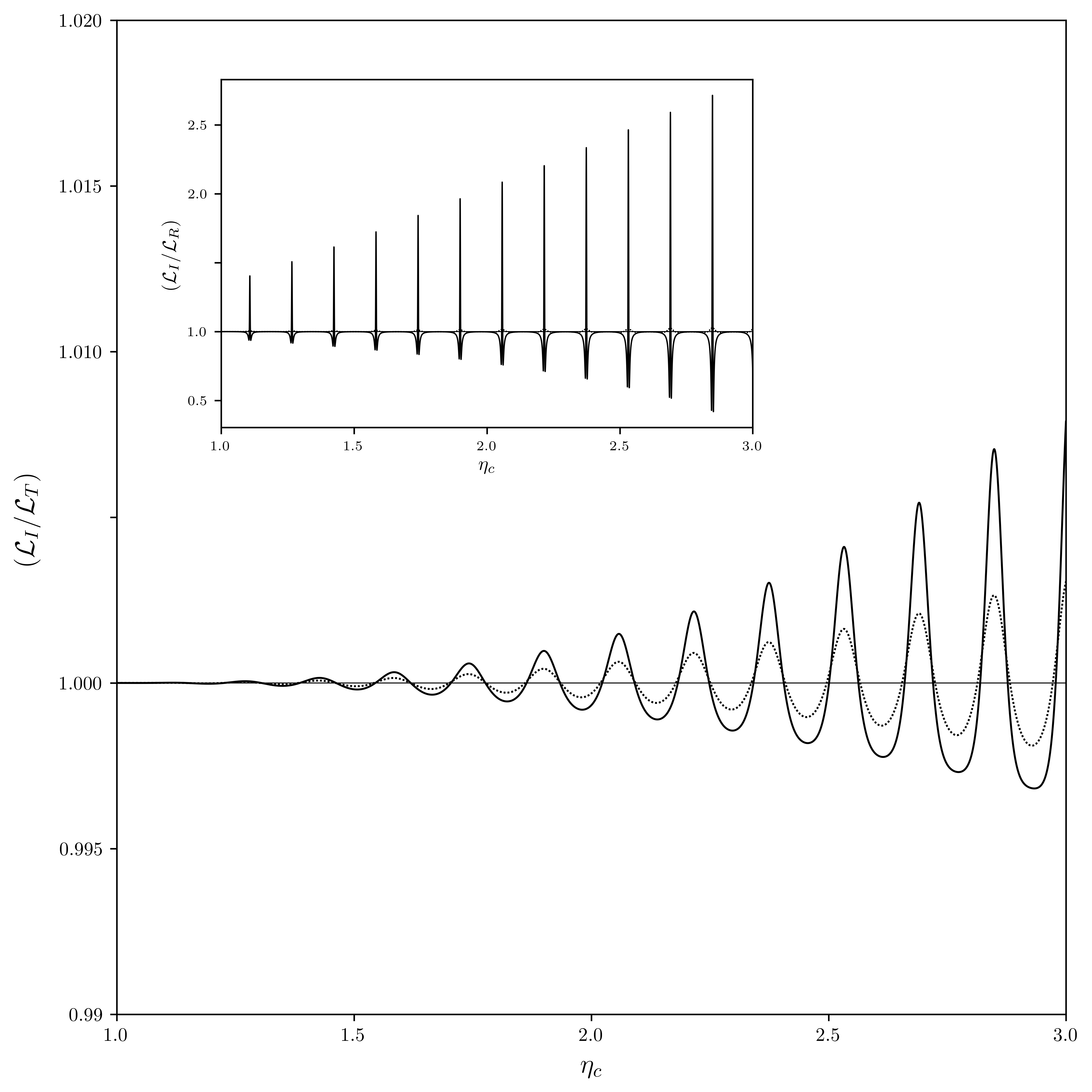}
  	\caption{
  		Change in relative mean-square spectral length,   $\mathcal{L}_{I}^2   / \mathcal{L}_{\Gamma}^2$, as a function of $\eta_c$ with same representative values of Fig.~\ref{fig:Relative_Delta_omega_vs_eta}, for transmitted pulse(main plot), and reflected one(inset plot) for two values of extinction coefficient, $\kappa_c$(solid for $\kappa_c$=0.002, and dotted for $\kappa_c$=0.02). A value greater than one indicates spectral broadening due to group velocity dispersion, while a value less than one indicates spectral narrowing. The reflected pulse exhibits more extreme distortion, as its spectrum is more sensitive to the Fabry-Perot resonances of the slab.}
  	\label{fig:Relative_Delta_L_vs_eta}
\end{figure}

\begin{figure}[!ht]
  	\includegraphics[width=\textwidth]{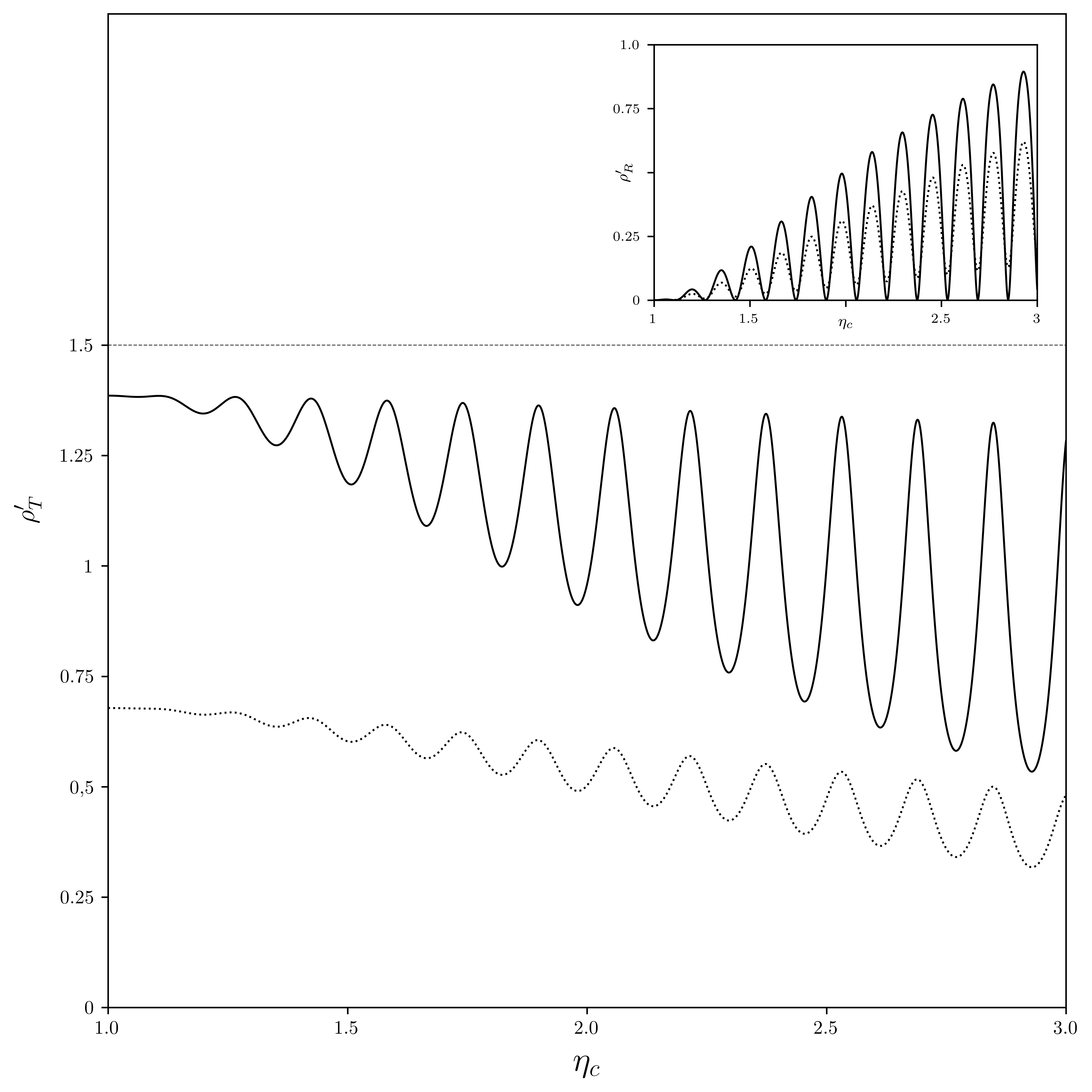}
  	\caption{Degradation of the effective peak squeezing parameter, $\rho_{\Gamma}^\prime / \rho_{I}$, for transmitted (main) and reflected (inset) continuum pulses  as a function of $\eta_c$. The representative values are the same as Fig.~\ref{fig:Relative_Delta_omega_vs_eta}. The ratio is always less than unity, confirming that absorption degrades non-classical squeezing. Degradation is more pronounced for higher loss (`$\kappa_c$=0.02`, dotted) and is maximized at Fabry-Perot resonances due to increased light-matter interaction.}
  	\label{fig:Effective_rho_Gamma_vs_eta}
\end{figure}

\begin{figure}[!ht]
  	\includegraphics[width=\textwidth]{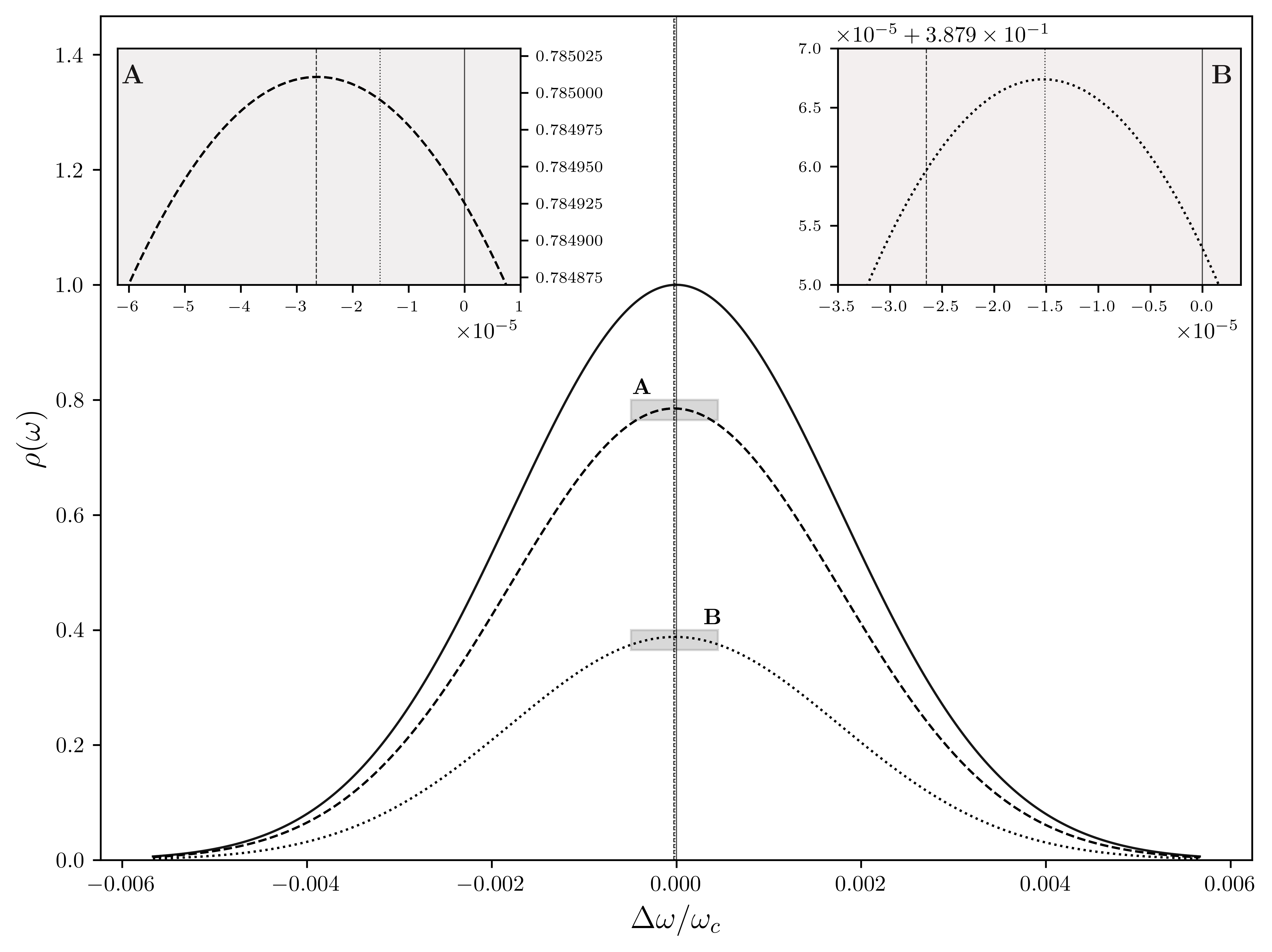}
  	\caption{Relative squeezing spectra, $\rho(\omega)$, as a function of normalized frequency shift,  $\left(\Delta\omega / \omega_c\right)$,  for incident pulse(solid), together with transmitted one for two values of $\kappa$(dashed for $\kappa_c$=0.002, and dotted for $\kappa_c$=0.02). Other representative values ar as the same as Fig.~\ref{fig:Relative_Delta_omega_vs_eta}. two narrow gray areas, \textbf{A} and \textbf{B}, zoomed in upper left and upper right of the main plot, show details of peak shifts in both transmitted pulses for two values of $\kappa$ respectively. The slab spectrally reshapes the pulse, introducing asymmetry and a small negative frequency shift of the peak (detailed in insets A and B). The overall reduction in `$\rho(\omega)$` across the spectrum visualizes the squeezing degradation caused by absorption.}
  	\label{fig:Rho-vs-Delta_omega}
\end{figure}

The numerical evaluation of these quantities reveals the rich structure of the scattering process. Figure 4 illustrates the relative frequency shift, $\Delta\omega_{\Gamma}/\omega_{c}$, as a function of the refractive index $\eta_{c}$ for two values of the extinction coefficient $\kappa$. The rapid oscillatory behavior arises from the interference of multiple internal reflections—the Fabry-Pérot resonances—within the slab. We observe that increased absorption (dotted line) dampens these resonances compared to the lower-loss case (solid line), as the internal cavity modes are more effectively suppressed by dissipation.

The spectral reshaping of the pulse is further detailed in Figure 5, which depicts the relative mean-square spectral length, $\mathcal{L}_{I}^{2}/\mathcal{L}_{\Gamma}^{2}$. Deviations from unity here indicate spectral broadening or narrowing induced by the medium's dispersion. Notably, the reflected pulse (inset) exhibits significantly larger distortions than the transmitted pulse, as the back-scattered field is highly sensitive to the exact resonance conditions of the slab interfaces.

The degradation of the quantum noise reduction is quantified in Figure 6, which plots the effective peak squeezing factor, $\rho_{\Gamma}^{\prime}/\rho_{I}$, against $\eta_c$. The ratio is strictly less than unity for all refractive indices, confirming that the slab acts as a decoherence channel that inevitably mixes vacuum noise with the squeezed state. This degradation is most pronounced at the Fabry-Pérot resonances, where the field spends the maximum amount of time interacting with the absorbing medium.

Finally, Figure 7 presents the full relative squeezing spectra, $\rho(\omega)$, as a function of the normalized frequency shift. This visualization explicitly demonstrates the asymmetry and the negative frequency shift of the peak discussed in Eq. (4.32). The overall reduction in the height of the spectral curve for the transmitted pulse, particularly for higher $\kappa$, visually confirms the irreversible loss of squeezing information across the pulse bandwidth.

\newpage

\subsection{Output Quadrature Noise Spectra}

The power spectrum analysis reveals how the energy distribution of the pulse is modified. However, to fully characterize the scattering of squeezed light, we must compute the noise in the field quadratures, generalizing the single-mode results of Sec. 3 to the continuum case.

We define the generalized quadrature operator for the output field  measured at time $t$ with a local oscillator phase $\phi$:
\begin{equation}
	\hat{X}^{\prime}_{\Gamma}(t, \phi) = \int_0^{\infty} d\omega\left[ \hat{a}_{\Gamma}(\omega) e^{-i(\omega t - \phi)} + \hat{a}_{\Gamma}^\dagger(\omega) e^{i(\omega t - \phi)} \right], \qquad \Gamma\in \{T,R\}.
\end{equation}
The variance of this operator, $\left(\Delta X^{\prime}_{\Gamma}(t, \phi)\right)^2 = \langle \left( \hat{X}^{\prime}_{\Gamma}(t, \phi) \right)^2 \rangle - \langle \hat{X}^{\prime}_{\Gamma}(t, \phi) \rangle^2$, quantifies the quantum noise at a specific point in time and for a chosen quadrature angle.

Substituting the input-output relations (2.10) and assuming the incident state is the squeezed vacuum (Eq. (4.1) with $\alpha(\omega)=0$), the calculation proceeds similarly to the single-mode case but involves convolution integrals due to the continuum of frequencies. Under the narrow-bandwidth approximation and assuming the pulse is long compared to the optical period, the variance simplifies to a form that closely mirrors the single-mode result:
\begin{equation}
	\left(\Delta X^{\prime}_{\Gamma}\right)^2 \approx \frac{1}{4} \left\{ 1 - |C_{\Gamma}(\omega_c)|^2 \left( 1 - e^{-2\rho_I} \right) + 2 \langle \hat{F}_{\Gamma}^\dagger \hat{F}_{\Gamma} \rangle \right\}
\end{equation}
for the optimally detected quadrature ($\phi = \theta + \delta_{\Gamma}$). Equation (4.39) demonstrates that the central conclusion from the single-mode analysis holds for continuum pulses: the dielectric slab selectively attenuates the incident squeezing, with the transmission (or reflection) coefficient dictating the final noise level. However, the continuum case introduces an additional layer of complexity due to the frequency dependence of these coefficients.

Since $T_S(\omega)$ varies across the pulse bandwidth, the degree of squeezing preservation is not uniform. The term $|C_\Gamma(\omega_c)|^2 (1 - e^{-2\rho_I})$ represents the geometric loss of the squeezed correlation due to the slab's boundaries and absorption. The final term, $2\langle \hat{F}_\Gamma^\dagger \hat{F}_\Gamma \rangle$, represents the "thermalization" of the field; even at zero temperature, the vacuum fluctuations associated with the slab's absorption add uncorrelated noise to the quadrature.

Consequently, Eq. (4.39) implies that the output pulse is generally no longer a minimum uncertainty state. The variance is inevitably pulled towards (or above) the Standard Quantum Limit (SQL) by the medium's dissipation. This degradation is most pronounced near the slab's resonant frequencies, where the light-matter interaction—and thus the coupling to the noise reservoir—is strongest. For high-precision applications relying on squeezed light, this necessitates a careful optimization of the slab's optical thickness to operate away from absorption peaks while maintaining high transmission.

\subsection{The average Poynting vector}
The average Poynting vector provides the space-time description of the energy flow and serves as a consistency check for the spectral power distribution derived in Section 4.1. We define the average normal-order Poynting vector as:
\begin{equation}
	\langle :\hat{S}(x,t): \rangle = \mu_0^{-1} \langle \hat{E}^-(x,t)\hat{B}^+(x,t) \rangle
\end{equation}
where the expectation value is taken over the product state $|\{\xi(\omega), \alpha(\omega)\}\rangle_R |0\rangle_L |F\rangle$. Using equations (2.7) and (2.9), the positive frequency parts of the field operators,  and the input-output relations, the total energy flux can be decomposed into three distinct contributions:
\begin{equation}
	\langle :\hat{S}_\Gamma(x,t): \rangle = \langle :\hat{S}_\Gamma(x,t): \rangle_{coh} + \langle :\hat{S}_\Gamma(x,t): \rangle_{sq} + \langle :\hat{S}_\Gamma(x,t): \rangle_{th}
\end{equation}
where $\Gamma \in \{T, R\}$ denotes the transmitted or reflected field.

The first term, $\langle :\hat{S}_\Gamma(x,t): \rangle_{coh}$, corresponds to the coherent amplitude of the pulse. It is formally identical to the result for coherent state scattering derived in our previous work, but now driven by the coherent amplitude $\alpha(\omega)$ of the squeezed state:
\begin{equation}
	\langle :\hat{S}_\Gamma(x,t): \rangle_{coh} = \left( \frac{\hbar}{4\pi\sigma} \right) \left| \int_{0}^{\infty} d\omega \sqrt{\omega} C_\Gamma(\omega) \alpha(\omega) e^{-i\omega(t - x/c)} \right|^2
\end{equation}
The second term, $\langle :\hat{S}_\Gamma(x,t): \rangle_{sq}$, arises purely from the non-classical squeezing correlations. Unlike a coherent state where $\langle \hat{a}^\dagger \hat{a} \rangle = |\alpha|^2$, the squeezed state carries an additional photon flux proportional to $\sinh^2[\rho(\omega)]$. Assuming the squeezing phase matches the propagation phase, this term represents a "quantum noise pulse" that travels alongside the coherent signal:
\begin{equation}
	\langle :\hat{S}_\Gamma(x,t): \rangle_{sq} = \left( \frac{\hbar}{4\pi\sigma} \right) \int_{0}^{\infty} d\omega \omega |C_\Gamma(\omega)|^2 \sinh^2[\rho_I(\omega)]
\end{equation}
The final term, $\langle :\hat{S}_\Gamma(x,t): \rangle_{th}$, is the thermal noise background contribution from the dissipative slab, which vanishes at zero temperature.

Integration of Eq. (4.41) over time yields the total energy density, confirming the Parseval-Plancherel relation with the power spectrum derived in Eq. (4.8):
\begin{equation}
\int_{-\infty}^{+\infty} dt \langle :\hat{S}_\Gamma(x,t): \rangle = \epsilon_0 c \int_{0}^{\infty} d\omega \mathcal{S}_\Gamma(\omega)
\end{equation}
To analyze the pulse reshaping analytically, we apply the narrow-bandwidth approximation to the coherent contribution (4.42). Assuming a Gaussian incident pulse $\alpha(\omega)$ as defined in Eq. (4.19) and the slab condition $\mathcal{L}_I \gg 2l\eta_c$, the transmitted and reflected energy pulses take the form:
\begin{equation}
\langle :\hat{S}_\Gamma(x,t): \rangle_{coh} \approx S_0 A_\Gamma \exp\left[ -\frac{2(x - ct + \Delta x_\Gamma)^2}{\mathcal{L}_\Gamma^2} \right]
\end{equation}
where $S_0$ is the peak intensity of the incident pulse. The spatial shift $\Delta x_\Gamma$ and the modified pulse length $\mathcal{L}_\Gamma$ are determined by the dispersive properties of the slab coefficients $C_\Gamma(\omega)$ evaluated at the carrier frequency $\omega_c$:
\begin{equation}
	\Delta x_\Gamma = \frac{\text{Re}(A B^*)}{\text{Re}(B)}, \quad \mathcal{L}_\Gamma^2 = \frac{|B|^2}{\text{Re}(B)}
\end{equation}
with $A = 2lc\gamma_\Gamma$ and $B = \mathcal{L}_I^2 + 32c^2l^2\beta_\Gamma^2$. These expressions show that the energy envelope of the squeezed coherent pulse suffers the same temporal delays and broadening (or compression) as a standard coherent pulse. However, the quantum noise properties carried by this packet are uniquely modified by the slab's absorption profile as detailed in Section 4.2. In the regime where the pulse length is comparable to or smaller than the optical thickness ($\mathcal{L}_I \ll 2l\eta_c$), the single Gaussian profile breaks down. The output decomposes into a train of resolved pulses arising from multiple internal reflections. The average Poynting vector then becomes a sum over these partial waves:
\begin{equation}
	\langle :\hat{S}_T(x,t): \rangle_{coh} = S_0 |t_{1c}t_{2c}|^2 \exp\left[\frac{-4\kappa_c \omega_c l}{c}\right] \sum_{m=0}^{\infty} |r_c|^{4m} \exp\left[-\frac{2(x - ct + \Delta x_m)^2}{\mathcal{L}_I^2}\right]
\end{equation}
where $\Delta x_m = -2l[1 - (2m+1)\eta_c]$ represents the delay of the $m$-th internal echo. This train of pulses carries the squeezed quantum noise in discrete packets, with each echo suffering progressively greater decoherence due to the increased path length through the absorbing medium.

\section{Conclusions}
We have developed a fully quantum mechanical theory for the scattering of squeezed light by a dissipative, dispersive dielectric slab. Building upon the Green-function scheme of electromagnetic field quantization, we extended previous results for coherent states to the domain of non-classical squeezed states. This formalism allows for a consistent treatment of noise, ensuring that the commutation relations of the field are preserved even in the presence of absorption.

Our analysis of single-mode scattering reveals that the slab acts as a phase-sensitive filter for quantum noise. We derived explicit analytical expressions for the variances of the transmitted and reflected quadratures (Eqs. 3.20–3.23). These results show that squeezing is only perfectly preserved in the limiting case of a lossless, perfectly matched slab. In the presence of dissipation, the slab mixes the incident squeezed fluctuations with the vacuum (or thermal) noise of the medium, inevitably degrading the squeezing level. We observed that this degradation is oscillatory with respect to the slab thickness, due to the interference of vacuum fluctuations undergoing multiple reflections within the slab boundaries. 

Generalizing to continuum-mode squeezed pulses, we examined the effects of dispersion and finite bandwidth. Using the narrow-bandwidth approximation, we showed that the transmitted and reflected pulses retain a Gaussian profile but suffer from spectral reshaping and temporal delay. Our analysis of the average Poynting vector confirmed that the energy envelope of the squeezed pulse undergoes the same traversal time shifts as a classical pulse, while carrying the modified quantum noise packets analyzed in the spectral domain. Specifically, we identified a relative frequency shift and a change in the pulse duration (spectral narrowing or broadening) governed by the complex transmission and reflection coefficients. 

Most significantly, we introduced an effective spectral squeezing parameter, $\rho'_\Gamma(\omega)$, which quantifies the surviving non-classicality of the scattered pulse. Our numerical results demonstrate that this parameter is always less than the incident squeezing parameter $\rho_I$, confirming that propagation through a dispersive absorbing medium is an irreversible process that erodes quantum correlations.

The expressions derived here provide a necessary theoretical foundation for utilizing squeezed light in realistic optical systems where material dispersion and loss cannot be neglected. Whether for quantum communication channels or high-precision interferometric sensing, understanding these limits is crucial for predicting the achievable signal-to-noise ratio in practical experiments involving dielectric media.

\end{document}